\documentclass[notitlepage]{article}
\usepackage{amsfonts}
\usepackage{amssymb}
\usepackage{amsmath}
\usepackage[twocolumn=false]{geometry}
\usepackage[dvips]{graphicx}
\usepackage{geometry}

\input{tcilatex}

\begin{document}

\date{}
\title{A generalized Ramanujan Master Theorem applied to the evaluation of
Feynman diagrams}
\author{$^{a}$Iv\'{a}n Gonz\'{a}lez\thanks{%
e-mail : ivan.gonzalez@usm.cl} , $^{b}$V. H. Moll\thanks{%
e-mail : vhm@math.tulane.edu} , $^{a}$Iv\'{a}n Schmidt\thanks{%
e-mail : ivan.schmidt@usm.cl}\bigskip \\
$^{a}$Departamento de F\'{\i}sica\\
Universidad T\'{e}cnica Federico Santa Mar\'{\i}a\\
and\\
Centro Cient\'{\i}fico-Tecnol\'{o}gico de Valparaiso (CCTVal), Valparaiso,
Chile.\bigskip \\
$^{b}$Department of Mathematics, Tulane University, New Orleans, LA 70118,
USA.}
\maketitle

\begin{abstract}
Ramanujan Master Theorem is a technique developed by the indian
mathematician S. Ramanujan to evaluate a class of definite integrals. This
technique is used here to calculate the values of integrals associated with
specific Feynman diagrams.
\end{abstract}

\bigskip

\bigskip

\bigskip

\bigskip

\textbf{PACS }: 11.25.Db; 12.38.Bx.

\bigskip

\textbf{Keywords }: Perturbation theory; Scalar integrals; Multiloop Feynman
diagrams; Schwinger parameters; Ramanujan's Master Theorem (RMT).

\vfill\newpage

\section{Introduction}

Precise experimental measurements in high energy physics require, in its
theoretical counterpart, the development of techniques for the evaluation of
analytic objects associated with the corresponding Feynman diagrams. These
techniques have lately emphasized the automatization of calculations of
multiscale, multiloop diagrams.

Modern numerical methods for the evaluation of Feynman diagrams benefit from
analytical techniques employed as preliminary work to detect the presence of
divergences. Recent advances include a method based on the Bernstein-Tkachov
theorem for the corrections of one and two loop diagrams \cite{2}, a method
based on sector-decomposition, as applied in \cite{3} to a variety of
diagrams and to the evaluation of integrals in phase space, and a third
example is the one developed in \cite{4}, which contains numerical
evaluations of two-point functions by differential equation methods.

New analytic methods include techniques to reduce Feynman diagrams to a
small number of scalar integrals, such as integration by parts \cite{9}, the
use of Lorentz invariance \cite{10} and the use of symmetry \cite{11}.
Methods for the evaluation of scalar integrals include expansion by regions
\cite{12}, Mellin-Barnes transforms \cite{13}, relations among integrals in
different dimensions \cite{14} and differential equations \cite{15}.

This paper contains examples of an alternative method for the evaluation of
some Feynman diagrams. The ideas are based on the classical Ramanujan Master
Theorem (RMT), which is a favorite technique employed by the well-known
mathematician to evaluate definite integrals. The theoretical aspects of
this method are presented in \cite{Hardy} and a collection of modern
examples is given in \cite{rmt}. The application of this technique has been
illustrated in \cite{IGo} with the evaluation of some multidimensional
integrals obtained by the Schwinger parametrization of Feynman diagrams. The
goal of the present work is to use this technique to evaluate integrals
associated to two and three loop diagrams. The method works for a large
variety of definite integrals and our first example illustrates this by
computing the Mellin transform of a Bessel function. The next example
illustrates the evaluation a multidimensional integral corresponding to the
massless bubble Feynman diagram. Then we provide a generalization of RMT to
multiple integrals, required for multiloop calculations. Further
applications will be described in future work.

\section{Ramanujan's Master Theorem (RMT) and its generalization : The
formalism}

Integrals of the form $\int_{0}^{\infty }dx\,x^{\nu -1}f(x)$ may be
evaluated by one of Ramanujan's favorite tools; the so-called Ramanujan
Master Theorem. It states that if $f(x)$ admits a series expansion of the
form

\begin{equation}
f(x)=\dsum\limits_{n=0}^{\infty }\varphi (n)\frac{(-x)^{n}}{n!}
\end{equation}%
in a neighborhood of $x=0$, with $f(0)=\varphi (0)\neq 0$, then

\begin{equation}
\dint\limits_{0}^{\infty }dx\;x^{\nu -1}f(x)=\Gamma (\nu )\varphi (-\nu ).
\label{eq1}
\end{equation}%
The integral is the Mellin transform of $f(x)$ and the term $\varphi (-\nu )$
requires an extension of the function $\varphi $, initially defined only for
$\nu \in \mathbb{N}$. Details on the natural unique extension of $\varphi $
are given in \cite{rmt}. Observe that, for $\nu >0$, the condition $\varphi
(0)\neq 0$ guarantees the convergence of the integral near $x=0$. The proof
of Ramanujan Master Theorem and the precise conditions for its application
appear in Hardy \cite{Hardy}. The reader will find in \cite{rmt} many other
examples.

\subsection{The Mellin transform of a Bessel function}

The first illustration of RMT involves an integral containing the classical
Bessel function, given in its hypergeometric form by

\begin{equation}
J_{\alpha }\left( \sqrt{x}\right) =\left( \dfrac{\sqrt{x}}{2}\right)
^{\alpha }\dfrac{1}{\Gamma \left( 1+\alpha \right) }\;_{0}F_{1}\left( \left.
\begin{array}{c}
- \\
1+\alpha%
\end{array}%
\right\vert -\tfrac{1}{4}x\right) .
\end{equation}%
Here $_{0}F_{1}$ is the hypergeometric function defined by

\begin{equation}
\;_{0}F_{1}\left( \left.
\begin{array}{c}
- \\
a%
\end{array}%
\right\vert x\right) =\dsum\limits_{n=0}^{\infty }\frac{1}{(a)_{n}}\frac{%
x^{n}}{n!},
\end{equation}%
using the Pochhammer symbol

\begin{equation}
(a)_{n}=\frac{\Gamma (a+n)}{\Gamma (a)}.
\end{equation}%
The integral to be evaluated here is

\begin{equation}
I=\dint\limits_{0}^{\infty }dx\;x^{\beta -1}J_{\alpha }\left( \sqrt{x}%
\right) ,  \label{inte}
\end{equation}%
and can be expressed as

\begin{equation}
\begin{array}{ll}
I & =\dint\limits_{0}^{\infty }dx\;x^{\beta -1}\left( \dfrac{\sqrt{x}}{2}%
\right) ^{\alpha }\dfrac{1}{\Gamma \left( 1+\alpha \right) }%
\tsum\limits_{n=0}^{\infty }\dfrac{\left( -1\right) ^{n}}{n!}\dfrac{1}{%
\left( 1+\alpha \right) _{n}}\dfrac{x^{n}}{4^{n}} \\
&  \\
& =\dint\limits_{0}^{\infty }\dfrac{dx}{x}\;\tsum\limits_{n=0}^{\infty }%
\tfrac{\left( -1\right) ^{n}}{n!}\left[ \dfrac{1}{2^{\alpha +2n}\,\Gamma
\left( 1+\alpha +n\right) }\right] x^{n+\beta +\tfrac{\alpha }{2}}.%
\end{array}%
\end{equation}%
The data matches the statement of RMT with

\begin{equation}
\varphi (n)=\frac{1}{2^{\alpha +2n}\,\Gamma (1+\alpha +n)}.
\end{equation}%
Therefore, the evaluation

\begin{equation}
I=\frac{\Gamma (-n^{\ast })}{2^{\alpha +2n^{\ast }}\Gamma (1+\alpha +n^{\ast
})}
\end{equation}%
is obtained directly from RMT. Here $n^{\ast }=-\left( \beta +\frac{\alpha }{%
2}\right) $. For information to be used in the latter sections, observe that
$n^{\ast }$ is the solution to

\begin{equation}
n+\beta +\tfrac{\alpha }{2}=0.
\end{equation}%
It follows that

\begin{equation}
\dint\limits_{0}^{\infty }dx\;x^{\beta -1}J_{\alpha }\left( \sqrt{x}\right)
=2^{2\beta }\dfrac{\Gamma \left( \beta +\tfrac{\alpha }{2}\right) }{\Gamma
\left( 1+\dfrac{\alpha }{2}-\beta \right) }.
\end{equation}%
This result appears as entry $6.561.14$ in the table of integrals \cite{gr}.

\subsection{A second example : the Feynman diagram of a bubble}

\begin{equation}
G=\begin{minipage}{5.9cm} \includegraphics[scale=.6] {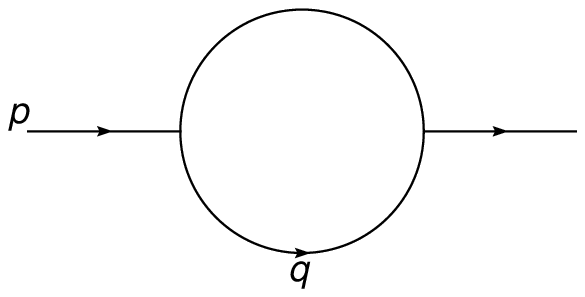}
\end{minipage}  \label{figb}
\end{equation}%
The next example illustrates the evaluation a multidimensional integral
corresponding to the massless bubble Feynman diagram shown in $\left( \ref%
{figb}\right) $. The result is well-known \cite{bubble}. In momentum space
the corresponding integral is given by

\begin{equation}
G=\dint \frac{d^{D}q}{i\pi ^{D/2}}\frac{1}{\left[ q^{2}\right] ^{a_{1}}\left[
\left( p-q\right) ^{2}\right] ^{a_{2}}},
\end{equation}%
where the parameters $\left\{ a_{i}\right\} $ are arbitrary. The Schwinger
representation corresponding to this diagram produces

\begin{equation}
G=\dfrac{\left( -1\right) ^{-\frac{D}{2}}}{\Gamma \left( a_{1}\right) \Gamma
\left( a_{2}\right) }\dint\limits_{0}^{\infty }\dint\limits_{0}^{\infty
}dx\,dy\;x^{a_{1}-1}y^{a_{2}-1}\;\frac{\exp \left( -\dfrac{xy}{x+y}%
\;p^{2}\right) }{\left( x+y\right) ^{\frac{D}{2}}}.
\end{equation}%
In order to apply RMT to evaluate this integral, each part of the integrand
is expanded in a Taylor series. In situations where options are available,
the optimal course of action seems to be to minimize the number of
expansions. At the moment this is at the heuristic level. In the present
example, it is convenient to first expand the exponential function

\begin{equation}
\exp \left( -\dfrac{xy}{x+y}\;p^{2}\right) =\sum_{n=0}^{\infty }\tfrac{%
\left( -1\right) ^{n}}{n!}\left( p^{2}\right) ^{n}\frac{x^{n}y^{n}}{\left(
x+y\right) ^{n}},
\end{equation}%
to produce the expression

\begin{equation}
G=\dfrac{\left( -1\right) ^{-\frac{D}{2}}}{\Gamma \left( a_{1}\right) \Gamma
\left( a_{2}\right) }\dint\limits_{0}^{\infty }\dint\limits_{0}^{\infty
}dx\,dy\;x^{a_{1}-1}y^{a_{2}-1}\;\tsum\limits_{n=0}^{\infty }\tfrac{\left(
-1\right) ^{n}}{n!}\left( p^{2}\right) ^{n}\frac{x^{n}y^{n}}{\left(
x+y\right) ^{\frac{D}{2}+n}}.  \label{G-int1}
\end{equation}%
The final step is to expand the binomial term $(x+y)^{-D/2-n}$ in the form

\begin{equation}
\frac{1}{\left( x+y\right) ^{\tfrac{D}{2}+n}}=\dsum\limits_{k=0}^{\infty }%
\tfrac{\left( -1\right) ^{k}}{k!}\left( \tfrac{D}{2}+n\right)
_{k}x^{-D/2-n-k}y^{k},
\end{equation}%
where the Pochhammer term is

\begin{equation}
\left( \tfrac{D}{2}+n\right) _{k}=\frac{\Gamma \left( \tfrac{D}{2}%
+n+k\right) }{\Gamma \left( \tfrac{D}{2}+n\right) }.
\end{equation}%
Replacing in \eqref{G-int1} gives

\begin{equation}
G=\dfrac{\left( -1\right) ^{-\frac{D}{2}}}{\Gamma \left( a_{1}\right) \Gamma
\left( a_{2}\right) }\dint\limits_{0}^{\infty }\dint\limits_{0}^{\infty }%
\frac{dx}{x}\frac{dy}{y}\;\tsum\limits_{k=0}^{\infty
}\tsum\limits_{n=0}^{\infty }\tfrac{\left( -1\right) ^{n}}{n!}\tfrac{\left(
-1\right) ^{k}}{k!}\left( p^{2}\right) ^{n}\left( \tfrac{D}{2}+n\right)
_{k}\;x^{-k+a_{1}-\frac{D}{2}}y^{k+n+a_{2}},
\end{equation}%
and $x\longrightarrow 1/x$ produces the alternative expression

\begin{equation}
G=\dfrac{\left( -1\right) ^{-\frac{D}{2}}}{\Gamma \left( a_{1}\right) \Gamma
\left( a_{2}\right) }\dint\limits_{0}^{\infty }\dint\limits_{0}^{\infty }%
\dfrac{dx}{x}\dfrac{dy}{y}\;\tsum\limits_{k=0}^{\infty
}\tsum\limits_{n=0}^{\infty }\tfrac{\left( -1\right) ^{n}}{n!}\tfrac{\left(
-1\right) ^{k}}{k!}\left( p^{2}\right) ^{n}\left( \tfrac{D}{2}+n\right)
_{k}\;x^{k-a_{1}+\frac{D}{2}}y^{k+n+a_{2}}.
\end{equation}%
There are several options to employ RMT to evaluate this integral :

\begin{itemize}
\item[a)] Evaluate the integral in the $x$-variable and the expansion in the
index $k$ :
\end{itemize}

\begin{equation*}
\dint\limits_{0}^{\infty }dx...\tsum\limits_{k=0}^{\infty }...\dfrac{\left(
-x\right) ^{k}}{k!}.
\end{equation*}

\begin{itemize}
\item[b)] Evaluate the integral in the $y$-variable jointly with the series
with index $k$ :
\end{itemize}

\begin{equation*}
\dint\limits_{0}^{\infty }dy...\tsum\limits_{k=0}^{\infty }...\dfrac{\left(
-y\right) ^{k}}{k!}.
\end{equation*}

\begin{itemize}
\item[c)] Evaluate the integral in the $y$-variable jointly with the series
with index $n$ :
\end{itemize}

\begin{equation*}
\dint\limits_{0}^{\infty }dy...\tsum\limits_{n=0}^{\infty }...\dfrac{\left(
-y\right) ^{n}}{n!}.
\end{equation*}%
It is now shown that each of these options produces the same result.

\subsection{Solution with option (a)}

The first procedure to evaluate $G$ is given by

\begin{equation}
G_{(a)}=\dfrac{\left( -1\right) ^{-\frac{D}{2}}}{\Gamma \left( a_{1}\right)
\Gamma \left( a_{2}\right) }\dint\limits_{0}^{\infty }\dfrac{dy}{y}\left[
\dint\limits_{0}^{\infty }\dfrac{dx}{x}\;\tsum\limits_{k=0}^{\infty }\dfrac{%
\left( -1\right) ^{k}}{k!}\varphi \left( k\right) \;x^{k-a_{1}+\frac{D}{2}}%
\right] ,  \label{eq2}
\end{equation}%
where $\varphi (k)$ is defined by

\begin{equation}
\varphi \left( k\right) =\dsum\limits_{n=0}^{\infty }\dfrac{\left( -1\right)
^{n}}{n!}\left( p^{2}\right) ^{n}\left( \tfrac{D}{2}+n\right)
_{k}\;y^{k+n+a_{2}}.
\end{equation}%
The value of \eqref{eq2} is given by RMT as

\begin{equation}
\begin{array}{ll}
G_{(a)} & =\dfrac{\left( -1\right) ^{-\frac{D}{2}}}{\Gamma \left(
a_{1}\right) \Gamma \left( a_{2}\right) }\dint\limits_{0}^{\infty }\dfrac{dy%
}{y}\;\Gamma (-k^{\ast })\varphi \left( k^{\ast }\right) ,\text{ with }%
k^{\ast }=-\frac{D}{2}+a_{1} \\
&  \\
& =\dfrac{\left( -1\right) ^{-\frac{D}{2}}}{\Gamma \left( a_{1}\right)
\Gamma \left( a_{2}\right) }\Gamma (\frac{D}{2}-a_{1})\dint\limits_{0}^{%
\infty }\dfrac{dy}{y}\;\tsum\limits_{n=0}^{\infty }\tfrac{\left( -1\right)
^{n}}{n!}\left( p^{2}\right) ^{n}\left( \frac{D}{2}+n\right) _{a_{1}-\frac{D%
}{2}}\;y^{a_{1}+a_{2}-\frac{D}{2}+n} \\
&  \\
& =\dfrac{\left( -1\right) ^{-\frac{D}{2}}}{\Gamma \left( a_{1}\right)
\Gamma \left( a_{2}\right) }\Gamma (\frac{D}{2}-a_{1})\dint\limits_{0}^{%
\infty }\dfrac{dy}{y}\;\tsum\limits_{n=0}^{\infty }\tfrac{\left( -1\right)
^{n}}{n!}\left( p^{2}\right) ^{n}\dfrac{\Gamma \left( a_{1}+n\right) }{%
\Gamma \left( \frac{D}{2}+n\right) }y^{n+a_{1}+a_{2}-\frac{D}{2}}.%
\end{array}%
\end{equation}%
RMT is applied one more time to the last integral to obtain

\begin{equation}
G_{(a)}=\dfrac{\left( -1\right) ^{-\frac{D}{2}}}{\Gamma \left( a_{1}\right)
\Gamma \left( a_{2}\right) }\Gamma (\tfrac{D}{2}-a_{1})\Gamma (-n^{\ast
})\;\left( p^{2}\right) ^{n^{\ast }}\dfrac{\Gamma \left( a_{1}+n^{\ast
}\right) }{\Gamma \left( \frac{D}{2}+n^{\ast }\right) },
\end{equation}%
with $n^{\ast }=-a_{1}-a_{2}+\frac{D}{2}$. Therefore, option (a) gives the
value of $G$ as

\begin{equation}
G_{(a)}=(-1)^{-\frac{D}{2}}\left( p^{2}\right) ^{\frac{D}{2}-a_{1}-a_{2}}%
\dfrac{\Gamma (a_{1}+a_{2}-\frac{D}{2})\Gamma (\frac{D}{2}-a_{1})\Gamma
\left( \frac{D}{2}-a_{2}\right) }{\Gamma (a_{1})\Gamma (a_{2})\Gamma \left(
D-a_{1}-a_{2}\right) }.
\end{equation}

\subsection{Solution with option (b)}

A similar argument now yields

\begin{equation}
G_{\left( b\right) }=\dfrac{\left( -1\right) ^{-\frac{D}{2}}}{\Gamma \left(
a_{1}\right) \Gamma \left( a_{2}\right) }\dint\limits_{0}^{\infty }\dfrac{dx%
}{x}\left[ \dint\limits_{0}^{\infty }\dfrac{dy}{y}\;\tsum\limits_{k=0}^{%
\infty }\tfrac{\left( -1\right) ^{k}}{k!}\varphi \left( k\right)
\;y^{k+n+a_{2}}\right]
\end{equation}%
with

\begin{equation}
\varphi \left( k\right) =\dsum\limits_{n=0}^{\infty }\tfrac{\left( -1\right)
^{n}}{n!}\left( p^{2}\right) ^{n}\left( \tfrac{D}{2}+n\right)
_{k}\;x^{k-a_{1}+\frac{D}{2}}.
\end{equation}%
Therefore

\begin{equation}
\begin{array}{ll}
G_{(b)} & =\dfrac{\left( -1\right) ^{-\frac{D}{2}}}{\Gamma \left(
a_{1}\right) \Gamma \left( a_{2}\right) }\dint\limits_{0}^{\infty }\dfrac{dx%
}{x}\;\Gamma (-k^{\ast })\varphi \left( k^{\ast }\right) ,\text{ \ \ with }%
k^{\ast }=-n-a_{2} \\
&  \\
& =\dfrac{\left( -1\right) ^{-\frac{D}{2}}}{\Gamma \left( a_{1}\right)
\Gamma \left( a_{2}\right) }\dint\limits_{0}^{\infty }\dfrac{dx}{x}%
\;\tsum\limits_{n=0}^{\infty }\tfrac{\left( -1\right) ^{n}}{n!}\Gamma
(n+a_{2})\left( p^{2}\right) ^{n}\left( \frac{D}{2}+n\right)
_{-n-a_{2}}\;x^{-n-a_{2}-a_{1}+\frac{D}{2}} \\
&  \\
& =\dfrac{\left( -1\right) ^{-\frac{D}{2}}}{\Gamma \left( a_{1}\right)
\Gamma \left( a_{2}\right) }\dint\limits_{0}^{\infty }\dfrac{dx}{x}%
\;\dsum\limits_{n=0}^{\infty }\tfrac{\left( -1\right) ^{n}}{n!}\left(
p^{2}\right) ^{n}\dfrac{\Gamma (n+a_{2})\Gamma \left( \frac{D}{2}%
-a_{2}\right) }{\Gamma \left( \frac{D}{2}+n\right) }\;x^{-n-a_{2}-a_{1}+%
\frac{D}{2}}.%
\end{array}%
\end{equation}%
The change of variables $x\mapsto 1/x$ gives

\begin{equation}
G_{(b)}=\dfrac{\left( -1\right) ^{-\frac{D}{2}}}{\Gamma \left( a_{1}\right)
\Gamma \left( a_{2}\right) }\dint\limits_{0}^{\infty }\dfrac{dx}{x}%
\;\tsum_{n=0}^{\infty }\tfrac{\left( -1\right) ^{n}}{n!}\left( p^{2}\right)
^{n}\dfrac{\Gamma (n+a_{2})\Gamma \left( \frac{D}{2}-a_{2}\right) }{\Gamma
\left( \frac{D}{2}+n\right) }\;x^{n+a_{2}+a_{1}-\frac{D}{2}},
\end{equation}%
and RMT produces

\begin{equation}
G_{(b)}=\dfrac{\left( -1\right) ^{-\frac{D}{2}}}{\Gamma \left( a_{1}\right)
\Gamma \left( a_{2}\right) }\Gamma (-n^{\ast })\;\left( p^{2}\right)
^{n^{\ast }}\dfrac{\Gamma (n^{\ast }+a_{2})\Gamma \left( \frac{D}{2}%
-a_{2}\right) }{\Gamma \left( \frac{D}{2}+n^{\ast }\right) },
\end{equation}%
with $n^{\ast }=-a_{2}-a_{1}+\frac{D}{2}$. Therefore, the value of $G$
obtained from option (b) is

\begin{equation}
G_{(b)}=\left( -1\right) ^{-\frac{D}{2}}\left( p^{2}\right) ^{\frac{D}{2}%
-a_{2}-a_{1}}\dfrac{\Gamma (\frac{D}{2}-a_{1})\Gamma \left( \frac{D}{2}%
-a_{2}\right) \Gamma (a_{2}+a_{1}-\tfrac{D}{2})}{\Gamma \left( a_{1}\right)
\Gamma \left( a_{2}\right) \Gamma \left( D-a_{2}-a_{1}\right) }.
\end{equation}

\subsection{Solution with option (c)}

Proceeding as before produces

\begin{equation}
G_{\left( c\right) }=\dfrac{\left( -1\right) ^{-\frac{D}{2}}}{\Gamma \left(
a_{1}\right) \Gamma \left( a_{2}\right) }\dint\limits_{0}^{\infty }\dfrac{dx%
}{x}\left[ \dint\limits_{0}^{\infty }\dfrac{dy}{y}\;\tsum\limits_{n=0}^{%
\infty }\tfrac{\left( -1\right) ^{n}}{n!}\varphi \left( n\right)
\;y^{k+n+a_{2}}\right]
\end{equation}%
with

\begin{equation}
\varphi \left( n\right) =\dsum\limits_{k=0}^{\infty }\tfrac{\left( -1\right)
^{k}}{k!}\left( p^{2}\right) ^{n}\left( \tfrac{D}{2}+n\right)
_{k}\;x^{k-a_{1}+\frac{D}{2}}.
\end{equation}%
This yields

\begin{equation}
G_{\left( c\right) }=\dfrac{\left( -1\right) ^{-\frac{D}{2}}}{\Gamma \left(
a_{1}\right) \Gamma \left( a_{2}\right) }\dint\limits_{0}^{\infty }\dfrac{dx%
}{x}\;\varphi \left( n^{\ast }\right) \Gamma (-n^{\ast })
\end{equation}%
with $n^{\ast }=-k-a_{2}$. Then

\begin{equation}
G_{\left( c\right) }=\dfrac{\left( -1\right) ^{-\frac{D}{2}}}{\Gamma \left(
a_{1}\right) \Gamma \left( a_{2}\right) }\dint\limits_{0}^{\infty }\dfrac{dx%
}{x}\;\tsum\limits_{k=0}^{\infty }\tfrac{\left( -1\right) ^{k}}{k!}\left(
p^{2}\right) ^{-k-a_{2}}\left( \tfrac{D}{2}-k-a_{2}\right) _{k}\;x^{k-a_{1}+%
\frac{D}{2}}\;\Gamma (k+a_{2})
\end{equation}%
with the equivalent form

\begin{equation}
G_{\left( c\right) }=\dfrac{\left( -1\right) ^{-\frac{D}{2}}}{\Gamma \left(
a_{1}\right) \Gamma \left( a_{2}\right) }\Gamma \left( \tfrac{D}{2}%
-a_{2}\right) \dint\limits_{0}^{\infty }\dfrac{dx}{x}\;\tsum\limits_{k=0}^{%
\infty }\tfrac{\left( -1\right) ^{k}}{k!}\left( p^{2}\right) ^{-k-a_{2}}%
\frac{\Gamma (k+a_{2})}{\Gamma \left( \tfrac{D}{2}-k-a_{2}\right) }%
\;x^{k-a_{1}+\frac{D}{2}}.
\end{equation}%
An application of RMT gives

\begin{equation}
G_{\left( c\right) }=\dfrac{\left( -1\right) ^{-\frac{D}{2}}}{\Gamma \left(
a_{1}\right) \Gamma \left( a_{2}\right) }\Gamma \left( \tfrac{D}{2}%
-a_{2}\right) \left( p^{2}\right) ^{-k^{\ast }-a_{2}}\frac{\Gamma \left(
-k^{\ast }\right) \Gamma (k^{\ast }+a_{2})}{\Gamma \left( \tfrac{D}{2}%
-k^{\ast }-a_{2}\right) }\text{ with }k^{\ast }=a_{1}-\tfrac{D}{2},
\end{equation}%
or equivalently

\begin{equation}
G_{\left( c\right) }=\left( -1\right) ^{-\frac{D}{2}}\left( p^{2}\right) ^{%
\frac{D}{2}-a_{1}-a_{2}}\frac{\Gamma \left( \tfrac{D}{2}-a_{2}\right) \Gamma
\left( \frac{D}{2}-a_{1}\right) \Gamma (a_{1}+a_{2}-\frac{D}{2})}{\Gamma
\left( a_{1}\right) \Gamma \left( a_{2}\right) \Gamma \left(
D-a_{1}-a_{2}\right) }.
\end{equation}%
Observe that each option produces the same value for the integral $G$; that
is, $G_{(a)}=G_{(b)}=G_{(c)}$.

\section{Generalization of RMT to multiple integrals. Multiloop calculations}

This section discusses a generalization of RMT to multidimensional integrals
of the form

\begin{equation}
I=\dint\limits_{0}^{\infty }dx_{1}\;x_{1}^{\nu
_{1}-1}\;...\dint\limits_{0}^{\infty }dx_{N}\;x_{N}^{\nu _{N}-1}\;f\left(
x_{1},...,x_{N}\right) .  \label{I}
\end{equation}%
As in the one-dimensional case, the function $f=f\left(
x_{1},...,x_{N}\right) $ is assumed to admit a Taylor expansion given by

\begin{equation}
\begin{array}{ll}
f\left( x_{1},...,x_{N}\right) = & \dsum\limits_{l_{1}=0}^{\infty
}...\dsum\limits_{l_{N}=0}^{\infty }\tfrac{\left( -1\right) ^{l_{1}}}{l_{1}!}%
...\tfrac{\left( -1\right) ^{l_{N}}}{l_{N}!}\;\varphi \left(
l_{1},...,l_{N}\right) \\
&  \\
& \times
\;x_{1}^{a_{11}l_{1}+...+a_{1N}l_{N}+b_{1}}...%
\;x_{N}^{a_{N1}l_{1}+...+a_{NN}l_{N}+b_{N}},%
\end{array}%
\end{equation}%
so that $I$ is expressed as

\begin{equation}
\begin{array}{ll}
I= & \dint\limits_{0}^{\infty }\dfrac{dx_{1}}{x_{1}}...\dint\limits_{0}^{%
\infty }\dfrac{dx_{N}}{x_{N}}\tsum\limits_{l_{1}=0}^{\infty
}...\tsum\limits_{l_{N}=0}^{\infty }\tfrac{\left( -1\right) ^{l_{1}}}{l_{1}!}%
...\tfrac{\left( -1\right) ^{l_{N}}}{l_{N}!}\;\varphi \left(
l_{1},..,l_{N}\right) \\
&  \\
& \times \;x_{1}^{a_{11}l_{1}+...+a_{1N}l_{N}+\widetilde{b}%
_{1}}..\;x_{N}^{a_{N1}l_{1}+...+a_{NN}l_{N}+\widetilde{b}_{N}}%
\end{array}
\label{eq3}
\end{equation}%
with $\widetilde{b}_{i}=\nu _{i}+b_{i}$ $\left( i=1,...,N\right) $.

Applying RMT in iterative manner produces

\begin{equation}
I=\dfrac{1}{\left\vert \det \left( \mathbf{A}\right) \right\vert }\;\Gamma
\left( -l_{1}^{\ast }\right) ...\Gamma \left( -l_{N}^{\ast }\right)
\;\varphi \left( l_{1}^{\ast },...,l_{N}^{\ast }\right)  \label{RMTG}
\end{equation}%
where $\det \left( \mathbf{A}\right) $ is the determinant

\begin{equation}
\det \left( \mathbf{A}\right) =\left\vert
\begin{array}{ccc}
a_{11} & \ldots & a_{1N} \\
\vdots & \ddots & \vdots \\
a_{N1} & \cdots & a_{NN}%
\end{array}%
\right\vert ,
\end{equation}%
and the variables $l_{i}^{\ast }\;\left( i=1,...,N\right) $ are solutions of
the linear system

\begin{equation}
\left\{
\begin{array}{ll}
a_{11}l_{1}+...+a_{1N}l_{N}+\widetilde{b}_{1} & =0 \\
\vdots & \vdots \\
a_{N1}l_{1}+...+a_{NN}l_{N}+\widetilde{b}_{N} & =0.%
\end{array}%
\right.
\end{equation}%
Details of the proof of this result appear in \cite{rmt}. This procedure
will be called the Generalized Ramanujan's Master Theorem (GRMT).

\section{Applications}

\subsection{Massive sunset diagram}

The integral evaluated first is associated to the diagram shown in the
figure. In momentum space, the integral is given by

\begin{equation}
G=\int \frac{d^{D}q_{1}}{i\pi ^{D/2}}\frac{d^{D}q_{2}}{i\pi ^{D/2}}\frac{1}{%
\left[ q^{2}-M^{2}\right] ^{a_{1}}}\frac{1}{\left[ \left( q_{1}-q_{2}\right)
^{2}\right] ^{a_{2}}}\frac{1}{\left[ (p+q_{2})^{2}\right] ^{a_{3}}}.
\end{equation}

{\ }%
\begin{equation}
G=\begin{minipage}{5.9cm} \includegraphics[scale=.6] {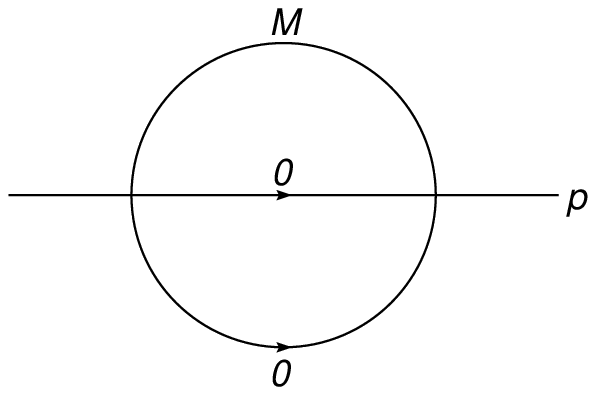}
\end{minipage}
\end{equation}%
In terms of the Schwinger parametrization, $G$ becomes

\begin{equation}
G=\dfrac{\left( -1\right) ^{-D}}{\Gamma \left( a_{1}\right) \Gamma \left(
a_{2}\right) \Gamma \left( a_{3}\right) }\dint\limits_{0}^{\infty
}\dint\limits_{0}^{\infty }\dint\limits_{0}^{\infty }d\overrightarrow{x}\;%
\dfrac{\exp \left( x_{1}M^{2}\right) \exp \left( -\dfrac{x_{1}x_{2}x_{3}}{%
x_{1}x_{2}+x_{1}x_{3}+x_{2}x_{3}}\;p^{2}\right) }{\left(
x_{1}x_{2}+x_{1}x_{3}+x_{2}x_{3}\right) ^{\frac{D}{2}}},
\end{equation}%
where $d\overrightarrow{x}=dx_{1}\,dx_{2}\,dx_{3}%
\;x_{1}^{a_{1}-1}x_{2}^{a_{2}-1}x_{3}^{a_{3}-1}$. In order to illustrate the
power of the GRMT method, the special physical case $p^{2}=M^{2}$ is
considered. The integral is now written as

\begin{equation}
G=\dfrac{\left( -1\right) ^{-D}}{\Gamma \left( a_{1}\right) \Gamma \left(
a_{2}\right) \Gamma \left( a_{3}\right) }\dint\limits_{0}^{\infty
}\dint\limits_{0}^{\infty }\dint\limits_{0}^{\infty }d\overrightarrow{x}\;%
\dfrac{\exp \left[ \dfrac{x_{1}^{2}\left( x_{2}+x_{3}\right) }{x_{1}\left(
x_{2}+x_{3}\right) +x_{2}x_{3}}\;M^{2}\right] }{\left[ x_{1}\left(
x_{2}+x_{3}\right) +x_{2}x_{3}\right] ^{\frac{D}{2}}}.
\end{equation}%
The expansion of the exponential function gives

\begin{equation}
\exp \left( \dfrac{x_{1}^{2}\left( x_{2}+x_{3}\right) }{x_{1}\left(
x_{2}+x_{3}\right) +x_{2}x_{3}}\;M^{2}\right) =\sum_{n_{1}=0}^{\infty }%
\tfrac{\left( -1\right) ^{n_{1}}}{n_{1}!}\left( -M^{2}\right) ^{n_{1}}\dfrac{%
x_{1}^{2n_{1}}\left( x_{2}+x_{3}\right) ^{n_{1}}}{\left[ x_{1}\left(
x_{2}+x_{3}\right) +x_{2}x_{3}\right] ^{n_{1}}}
\end{equation}%
and this produces

\begin{equation}
G=\dfrac{\left( -1\right) ^{-D}}{\Gamma \left( a_{1}\right) \Gamma \left(
a_{2}\right) \Gamma \left( a_{3}\right) }\dint\limits_{0}^{\infty
}\dint\limits_{0}^{\infty }\dint\limits_{0}^{\infty }d\overrightarrow{x}%
\;\tsum\limits_{n_{1}=0}^{\infty }\tfrac{\left( -1\right) ^{n_{1}}}{n_{1}!}%
\left( -M^{2}\right) ^{n_{1}}\dfrac{x_{1}^{2n_{1}}\left( x_{2}+x_{3}\right)
^{n_{1}}}{\left[ x_{1}\left( x_{2}+x_{3}\right) +x_{2}x_{3}\right] ^{\frac{D%
}{2}+n_{1}}}.  \label{eq4}
\end{equation}%
The binomial theorem is employed next to expand the integrand. This can be
done via the hypergeometric representation

\begin{equation}
\left( 1+x\right) ^{a}=\;_{1}F_{0}\left( \left.
\begin{array}{c}
-a \\
-%
\end{array}%
\right\vert -x\right) ,
\end{equation}%
or by the explicit formula

\begin{equation}
\left( x+y\right) ^{a}=\sum_{n=0}^{\infty }\tfrac{\left( -1\right) ^{n}}{n!}%
\frac{\Gamma \left( -a+n\right) }{\Gamma \left( -a\right) }\;x^{a-n}y^{n}.
\end{equation}%
The result is

\begin{equation}
\begin{array}{ll}
\dfrac{1}{\left[ x_{1}\left( x_{2}+x_{3}\right) +x_{2}x_{3}\right] ^{\frac{D%
}{2}+n_{1}}}= & \dsum\limits_{n_{2}=0}^{\infty }\tfrac{\left( -1\right)
^{n_{2}}}{n_{2}!}\dfrac{\Gamma \left( \frac{D}{2}+n_{1}+n_{2}\right) }{%
\Gamma \left( \frac{D}{2}+n_{1}\right) } \\
&  \\
& \times \;x_{1}^{-\frac{D}{2}-n_{1}-n_{2}}\left( x_{2}+x_{3}\right) ^{-%
\frac{D}{2}-n_{1}-n_{2}}x_{2}^{n_{2}}x_{3}^{n_{2}},%
\end{array}%
\end{equation}%
and \eqref{eq4} is now written as

\begin{equation}
\begin{array}{ll}
G= & \dfrac{\left( -1\right) ^{-D}}{\Gamma \left( a_{1}\right) \Gamma \left(
a_{2}\right) \Gamma \left( a_{3}\right) }\dint\limits_{0}^{\infty
}\dint\limits_{0}^{\infty }\dint\limits_{0}^{\infty }d\overrightarrow{x}%
\;\tsum\limits_{n_{1}=0}^{\infty }\tsum\limits_{n_{2}=0}^{\infty }\tfrac{%
\left( -1\right) ^{n_{1}}}{n_{1}!}\tfrac{\left( -1\right) ^{n_{2}}}{n_{2}!}%
\left( M^{2}\right) ^{n_{1}} \\
&  \\
& \times \;\dfrac{\Gamma \left( \frac{D}{2}+n_{1}+n_{2}\right) }{\Gamma
\left( \frac{D}{2}+n_{1}\right) }\;x_{1}^{n_{1}-\frac{D}{2}%
-n_{2}}x_{2}^{n_{2}}x_{3}^{n_{2}}\left( x_{2}+x_{3}\right) ^{-\frac{D}{2}%
-n_{2}}.%
\end{array}%
\end{equation}%
Only the binomial $(x_{2}+x_{3})^{-\frac{D}{2}-n_{2}}$ needs to be expanded.
This is done as before to produce

\begin{equation}
\left( x_{2}+x_{3}\right) ^{-\frac{D}{2}-n_{2}}=\sum_{n_{3}=0}^{\infty }%
\tfrac{\left( -1\right) ^{n_{3}}}{n_{3}!}\dfrac{\Gamma \left( \frac{D}{2}%
+n_{2}+n_{3}\right) }{\Gamma \left( \frac{D}{2}+n_{2}\right) }x_{2}^{-\frac{D%
}{2}-n_{2}-n_{3}}x_{3}^{n_{3}}.
\end{equation}%
The point has been reached for a direct application of GRMT to evaluate

\begin{equation}
\begin{array}{ll}
G= & \dint\limits_{0}^{\infty }\dint\limits_{0}^{\infty
}\dint\limits_{0}^{\infty }\dfrac{dx_{1}}{x_{1}}\dfrac{dx_{2}}{x_{2}}\dfrac{%
dx_{3}}{x_{3}} \\
&  \\
& \times \;\tsum\limits_{n_{1}=0}^{\infty }\tsum\limits_{n_{2}=0}^{\infty
}\tsum\limits_{n_{3}=0}^{\infty }\tfrac{\left( -1\right) ^{n_{1}}}{n_{1}!}%
\tfrac{\left( -1\right) ^{n_{2}}}{n_{2}!}\tfrac{\left( -1\right) ^{n_{3}}}{%
n_{3}!}x_{1}^{a_{1}-\frac{D}{2}+n_{1}-n_{2}}x_{2}^{a_{2}-\frac{D}{2}%
-n_{3}}x_{3}^{a_{3}+n_{2}+n_{3}} \\
&  \\
& \times \;\dfrac{\left( -1\right) ^{-D}}{\Gamma \left( a_{1}\right) \Gamma
\left( a_{2}\right) \Gamma \left( a_{3}\right) }\dfrac{\Gamma \left( \frac{D%
}{2}+n_{1}+n_{2}\right) }{\Gamma \left( \frac{D}{2}+n_{1}\right) }\dfrac{%
\Gamma \left( \frac{D}{2}+n_{2}+n_{3}\right) }{\Gamma \left( \frac{D}{2}%
+n_{2}\right) }\left( -M^{2}\right) ^{n_{1}}.%
\end{array}
\label{eqn5}
\end{equation}%
It follows that the integral $G$ is given by

\begin{equation}
G=\dfrac{\left( -1\right) ^{-D}}{\Gamma \left( a_{1}\right) \Gamma \left(
a_{2}\right) \Gamma \left( a_{3}\right) }\Gamma \left( -n_{1}^{\ast }\right)
\Gamma \left( -n_{2}^{\ast }\right) \Gamma \left( -n_{3}^{\ast }\right)
\dfrac{\Gamma \left( \frac{D}{2}+n_{1}^{\ast }+n_{2}^{\ast }\right) }{\Gamma
\left( \frac{D}{2}+n_{1}^{\ast }\right) }\dfrac{\Gamma \left( \frac{D}{2}%
+n_{2}^{\ast }+n_{3}^{\ast }\right) }{\Gamma \left( \frac{D}{2}+n_{2}^{\ast
}\right) }\left( -M^{2}\right) ^{n_{1}^{\ast }},  \label{G-value}
\end{equation}%
where the values assigned to the indices $\left\{ n_{i}^{\ast }\right\} $
are the unique solution to the linear system obtained from \eqref{eqn5}. In
detail,

\begin{equation}
\left\{
\begin{array}{l}
n_{1}-n_{2}=-a_{1}+\frac{D}{2}, \\
\\
n_{3}=a_{2}-\frac{D}{2}, \\
\\
n_{2}+n_{3}=-a_{3},%
\end{array}%
\right.
\end{equation}%
with solution

\begin{equation}
\left\{
\begin{array}{l}
n_{1}^{\ast }=-a_{1}-a_{2}-a_{3}+D, \\
\\
n_{2}^{\ast }=-a_{2}-a_{3}+\frac{D}{2}, \\
\\
n_{3}^{\ast }=-\frac{D}{2}+a_{2}.%
\end{array}%
\right.
\end{equation}%
Replacing in \eqref{G-value} yields

\begin{equation}
\begin{array}{ll}
G= & \left( -1\right) ^{-D}\dfrac{\Gamma \left( a_{1}+a_{2}+a_{3}-D\right)
\Gamma \left( a_{2}+a_{3}-\tfrac{D}{2}\right) \Gamma \left( \tfrac{D}{2}%
-a_{2}\right) }{\Gamma \left( a_{1}\right) \Gamma \left( a_{2}\right) \Gamma
\left( a_{3}\right) } \\
&  \\
& \times \;\dfrac{\Gamma \left( \frac{D}{2}-a_{3}\right) \Gamma \left(
2D-a_{1}-2a_{2}-2a_{3}\right) }{\Gamma \left( \frac{3D}{2}%
-a_{1}-a_{2}-a_{3}\right) \Gamma \left( D-a_{2}-a_{3}\right) }\left(
-M^{2}\right) ^{D-a_{1}-a_{2}-a_{3}}.%
\end{array}%
\end{equation}

\subsection{\textbf{Massless three loops ladder diagram}}

The next example gives the evaluation of the integral associated to the
diagram seen in Eq. $\left( \ref{eq6}\right) $. To illustrate the method in
a relatively simple situation, the conditions $P_{i}^{2}=0$ for $1\leq i\leq
4$ and $s=0$ are imposed.

{\ }%
\begin{equation}
G=\begin{minipage}{5.9cm} \includegraphics[scale=.7] {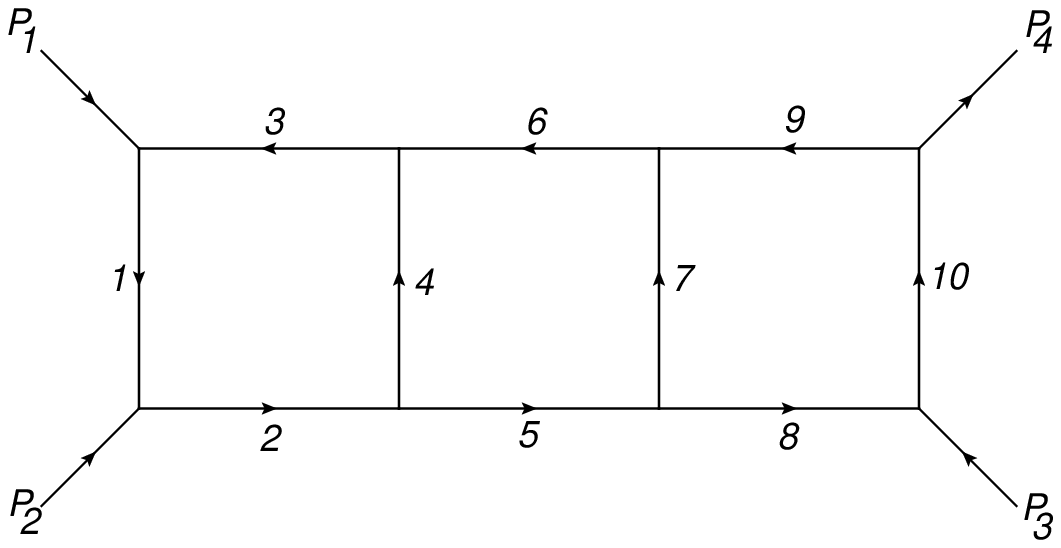}
\end{minipage}  \label{eq6}
\end{equation}%
The parametric representation of this diagram is given by the integral

\begin{equation}
G=\dfrac{\left( -1\right) ^{-\frac{3D}{2}}}{\Gamma \left( a_{1}\right)
...\Gamma \left( a_{10}\right) }\dint\limits_{0}^{\infty
}...\dint\limits_{0}^{\infty }d\overrightarrow{x}\;\dfrac{\exp \left( -%
\dfrac{x_{1}x_{4}x_{7}x_{10}}{U}\;t\right) }{U^{\frac{D}{2}}},
\end{equation}%
where $d\overrightarrow{x}=\prod_{j=1}^{10}dx_{j}\;x_{j}^{a_{j}-1}$ and the
polynomial $U$, written in a form adapted to the application of GRMT, is
given by

\begin{equation}
U=x_{5}\left( x_{7}+\mathbf{f}_{1}\right) \left( \mathbf{f}_{2}+x_{4}\right)
+x_{6}\left( x_{7}+\mathbf{f}_{1}\right) \left( \mathbf{f}_{2}+x_{4}\right)
+x_{4}\left( x_{7}+\mathbf{f}_{1}\right) \mathbf{f}_{2}+x_{7}\left( \mathbf{f%
}_{2}+x_{4}\right) \mathbf{f}_{1},
\end{equation}%
with

\begin{equation}
\left\{
\begin{array}{l}
\mathbf{f}_{1}=x_{8}+x_{9}+x_{10}, \\
\\
\mathbf{f}_{2}=x_{1}+x_{2}+x_{3}.%
\end{array}%
\right.
\end{equation}%
Expanding the exponential term produces

\begin{equation}
G=\dfrac{\left( -1\right) ^{-\frac{3D}{2}}}{\Gamma \left( a_{1}\right)
...\Gamma \left( a_{10}\right) }\dint\limits_{0}^{\infty
}...\dint\limits_{0}^{\infty }d\overrightarrow{x}\;\tsum\limits_{n_{1}=0}^{%
\infty }\dfrac{\left( -1\right) ^{n_{1}}}{n_{1}!}\left( t\right) ^{n_{1}}%
\dfrac{x_{1}^{n_{1}}x_{4}^{n_{1}}x_{7}^{n_{1}}x_{10}^{n_{1}}}{U^{\frac{D}{2}%
+n_{1}}},
\end{equation}%
and the polynomial $U$ is expanded using the multinomial theorem

\begin{equation}
\begin{array}{ll}
\left( x_{1}+...+x_{k-1}+x_{k}\right) ^{a}= & \dsum\limits_{n_{1}=0}^{\infty
}...\dsum\limits_{n_{k-1}=0}^{\infty }\dfrac{1}{n_{1}!...n_{k-1}!}\dfrac{%
\Gamma \left( 1+a\right) }{\Gamma \left( 1+a-n_{1}-...-n_{k-1}\right) } \\
&  \\
& \times \;x_{1}^{n_{1}}...x_{k-1}^{n_{k-1}}x_{k}^{a-n_{1}-...-n_{k-1}},%
\end{array}%
\end{equation}%
written in a form adapted to GRMT

\begin{equation}
\begin{array}{ll}
\left( x_{1}+...+x_{k-1}+x_{k}\right) ^{a}= & \dsum\limits_{n_{1}=0}^{\infty
}...\dsum\limits_{n_{k-1}=0}^{\infty }\tfrac{\left( -1\right) ^{n_{1}}}{%
n_{1}!}...\tfrac{\left( -1\right) ^{n_{k-1}}}{n_{k-1}!}\dfrac{\Gamma \left(
-a+n_{1}+...+n_{k-1}\right) }{\Gamma \left( -a\right) } \\
&  \\
& \times \;x_{1}^{n_{1}}...x_{k-1}^{n_{k-1}}x_{k}^{a-n_{1}-...-n_{k-1}}.%
\end{array}%
\end{equation}%
These expansions produce

\begin{equation}
\begin{array}{ll}
U^{-\frac{D}{2}-n_{1}}= & \dsum\limits_{n_{2}=0}^{\infty
}\dsum\limits_{n_{3}=0}^{\infty }\dsum\limits_{n_{4}=0}^{\infty }\tfrac{%
\left( -1\right) ^{n_{2}}}{n_{2}!}\tfrac{\left( -1\right) ^{n_{3}}}{n_{3}!}%
\tfrac{\left( -1\right) ^{n_{4}}}{n_{4}!}\dfrac{\Gamma \left( \frac{D}{2}%
+n_{1}+n_{2}+n_{3}+n_{4}\right) }{\Gamma \left( \frac{D}{2}+n_{1}\right) }
\\
&  \\
& \times \;\mathbf{f}_{1}^{-\frac{D}{2}-n_{1}-n_{2}-n_{3}-n_{4}}\mathbf{f}%
_{2}^{n_{2}}\left( x_{7}+\mathbf{f}_{1}\right) ^{n_{2}+n_{3}+n_{4}}\left(
x_{4}+\mathbf{f}_{2}\right) ^{-\frac{D}{2}-n_{1}-n_{2}} \\
&  \\
& \times \;x_{4}^{n_{2}}x_{5}^{n_{3}}x_{6}^{n_{4}}x_{7}^{-\frac{D}{2}%
-n_{1}-n_{2}-n_{3}-n_{4}}.%
\end{array}%
\end{equation}%
Similarly,

\begin{equation}
\left( x_{7}+\mathbf{f}_{1}\right)
^{n_{2}+n_{3}+n_{4}}=\sum_{n_{5}=0}^{\infty }\tfrac{\left( -1\right) ^{n_{5}}%
}{n_{5}!}\dfrac{\Gamma \left( -n_{2}-n_{3}-n_{4}+n_{5}\right) }{\Gamma
\left( -n_{2}-n_{3}-n_{4}\right) }x_{7}^{n_{5}}\mathbf{f}%
_{1}^{n_{2}+n_{3}+n_{4}-n_{5}},
\end{equation}%
and

\begin{equation}
\left( x_{4}+\mathbf{f}_{2}\right) ^{-\frac{D}{2}-n_{1}-n_{2}}=%
\sum_{n_{6}=0}^{\infty }\tfrac{\left( -1\right) ^{n_{6}}}{n_{6}!}\dfrac{%
\Gamma \left( \frac{D}{2}+n_{1}+n_{2}+n_{6}\right) }{\Gamma \left( \frac{D}{2%
}+n_{1}+n_{2}\right) }x_{4}^{n_{6}}\mathbf{f}_{2}^{-\frac{D}{2}%
-n_{1}-n_{2}-n_{6}}.
\end{equation}%
The result is
\begin{equation}
\begin{array}{ll}
G= & \dfrac{\left( -1\right) ^{-\frac{3D}{2}}}{\Gamma \left( a_{1}\right)
...\Gamma \left( a_{10}\right) }\dint\limits_{0}^{\infty
}...\dint\limits_{0}^{\infty }d\overrightarrow{x}\;\tsum\limits_{n_{1}=0}^{%
\infty }...\tsum\limits_{n_{6}=0}^{\infty }\tfrac{\left( -1\right) ^{n_{1}}}{%
n_{1}!}...\tfrac{\left( -1\right) ^{n_{6}}}{n_{6}!}\left( t\right) ^{n_{1}}
\\
&  \\
& \times \;\dfrac{\Gamma \left( \frac{D}{2}+n_{1}+n_{2}+n_{3}+n_{4}\right) }{%
\Gamma \left( \frac{D}{2}+n_{1}\right) }\dfrac{\Gamma \left(
-n_{2}-n_{3}-n_{4}+n_{5}\right) }{\Gamma \left( -n_{2}-n_{3}-n_{4}\right) }%
\dfrac{\Gamma \left( \frac{D}{2}+n_{1}+n_{2}+n_{6}\right) }{\Gamma \left(
\frac{D}{2}+n_{1}+n_{2}\right) } \\
&  \\
& \times
\;x_{1}^{n_{1}}x_{4}^{n_{1}+n_{2}+n_{6}}x_{5}^{n_{3}}x_{6}^{n_{4}}x_{7}^{-%
\frac{D}{2}-n_{2}-n_{3}-n_{4}+n_{5}}x_{10}^{n_{1}}\mathbf{f}_{1}^{-\frac{D}{2%
}-n_{1}-n_{5}}\mathbf{f}_{2}^{-\frac{D}{2}-n_{1}-n_{6}}.%
\end{array}%
\end{equation}%
Finally, the powers of $\mathbf{f}_{1}$ and $\mathbf{f}_{2}$ are expanded in
the form

\begin{equation}
\begin{array}{ll}
\mathbf{f}_{1}^{-\frac{D}{2}-n_{1}-n_{5}}= & \dsum\limits_{n_{7}=0}^{\infty
}\dsum\limits_{n_{8}=0}^{\infty }\dfrac{\left( -1\right) ^{n_{7}}}{n_{7}!}%
\dfrac{\left( -1\right) ^{n_{8}}}{n_{9}!}\dfrac{\Gamma \left( \frac{D}{2}%
+n_{1}+n_{5}+n_{7}+n_{8}\right) }{\Gamma \left( \frac{D}{2}%
+n_{1}+n_{5}\right) } \\
&  \\
& \times \;x_{8}^{n_{7}}x_{9}^{n_{8}}x_{10}^{-\frac{D}{2}%
-n_{1}-n_{5}-n_{7}-n_{8}},%
\end{array}%
\end{equation}%
and

\begin{equation}
\begin{array}{ll}
\mathbf{f}_{2}^{-\frac{D}{2}-n_{1}-n_{6}}= & \dsum\limits_{n_{9}=0}^{\infty
}\dsum\limits_{n_{10}=0}^{\infty }\tfrac{\left( -1\right) ^{n_{9}}}{n_{9}!}%
\tfrac{\left( -1\right) ^{n_{10}}}{n_{10}!}\dfrac{\Gamma \left( \frac{D}{2}%
+n_{1}+n_{6}+n_{9}+n_{10}\right) }{\Gamma \left( \frac{D}{2}%
+n_{1}+n_{6}\right) } \\
&  \\
& \times \;x_{1}^{n_{9}}x_{2}^{n_{10}}x_{3}^{-\frac{D}{2}%
-n_{1}-n_{6}-n_{9}-n_{10}}.%
\end{array}%
\end{equation}%
At this point the expression for $G$ is in the form required to apply GRMT :

\begin{equation}
\begin{array}{ll}
G= & \dfrac{\left( -1\right) ^{-\frac{3D}{2}}}{\Gamma \left( a_{1}\right)
...\Gamma \left( a_{10}\right) }\dint\limits_{0}^{\infty }\dfrac{dx_{1}}{%
x_{1}}...\dint\limits_{0}^{\infty }\dfrac{dx_{10}}{x_{10}}%
\;\tsum\limits_{n_{1}=0}^{\infty }...\tsum\limits_{n_{10}=0}^{\infty }\tfrac{%
\left( -1\right) ^{n_{1}}}{n_{1}!}...\tfrac{\left( -1\right) ^{n_{10}}}{%
n_{10}!}\;\varphi \left( n_{1},...,n_{10}\right) \\
&  \\
& \times \;x_{1}^{a_{1}+n_{1}+n_{9}}x_{2}^{a_{2}+n_{10}}x_{3}^{a_{3}-\frac{D%
}{2}%
-n_{1}-n_{6}-n_{9}-n_{10}}x_{4}^{a_{4}+n_{1}+n_{2}+n_{6}}x_{5}^{a_{5}+n_{3}}x_{6}^{a_{6}+n_{4}}x_{7}^{a_{7}-%
\frac{D}{2}-n_{2}-n_{3}-n_{4}+n_{5}} \\
&  \\
& \times \;x_{8}^{a_{8}+n_{7}}x_{9}^{a_{9}+n_{8}}x_{10}^{a_{10}-\frac{D}{2}%
-n_{5}-n_{7}-n_{8}},%
\end{array}%
\end{equation}%
with the notation

\begin{equation}
\begin{array}{ll}
\varphi \left( n_{1},...,n_{10}\right) = & \dfrac{\Gamma \left( \frac{D}{2}%
+n_{1}+n_{2}+n_{3}+n_{4}\right) }{\Gamma \left( \frac{D}{2}+n_{1}\right) }%
\dfrac{\Gamma \left( -n_{2}-n_{3}-n_{4}+n_{5}\right) }{\Gamma \left(
-n_{2}-n_{3}-n_{4}\right) }\dfrac{\Gamma \left( \frac{D}{2}%
+n_{1}+n_{2}+n_{6}\right) }{\Gamma \left( \frac{D}{2}+n_{1}+n_{2}\right) }
\\
&  \\
& \times \;\dfrac{\Gamma \left( \frac{D}{2}+n_{1}+n_{5}+n_{7}+n_{8}\right) }{%
\Gamma \left( \frac{D}{2}+n_{1}+n_{5}\right) }\dfrac{\Gamma \left( \frac{D}{2%
}+n_{1}+n_{6}+n_{9}+n_{10}\right) }{\Gamma \left( \frac{D}{2}%
+n_{1}+n_{6}\right) }\left( t\right) ^{n_{1}}.%
\end{array}
\notag
\end{equation}%
The value of $G$ is now obtained as a direct application of GRMT as

\begin{equation}
G=\dfrac{\left( -1\right) ^{-\frac{3D}{2}}}{\Gamma \left( a_{1}\right)
...\Gamma \left( a_{10}\right) }\dfrac{1}{\left\vert \det \left( \mathbf{A}%
\right) \right\vert }\;\varphi \left( n_{1}^{\ast },...,n_{10}^{\ast
}\right) \prod_{j=1}^{10}\Gamma \left( -n_{j}^{\ast }\right)
\end{equation}%
where the variables $\left\{ n_{i}^{\ast }\right\} $ are solutions of the
linear system

\begin{equation}
\left(
\begin{array}{cccccccccc}
1 & 0 & 0 & 0 & 0 & 0 & 0 & 0 & 1 & 0 \\
0 & 0 & 0 & 0 & 0 & 0 & 0 & 0 & 0 & 1 \\
-1 & 0 & 0 & 0 & 0 & -1 & 0 & 0 & -1 & -1 \\
1 & 1 & 0 & 0 & 0 & 1 & 0 & 0 & 0 & 0 \\
0 & 0 & 1 & 0 & 0 & 0 & 0 & 0 & 0 & 0 \\
0 & 0 & 0 & 1 & 0 & 0 & 0 & 0 & 0 & 0 \\
0 & -1 & -1 & -1 & 1 & 0 & 0 & 0 & 0 & 0 \\
0 & 0 & 0 & 0 & 0 & 0 & 1 & 0 & 0 & 0 \\
0 & 0 & 0 & 0 & 0 & 0 & 0 & 1 & 0 & 0 \\
0 & 0 & 0 & 0 & -1 & 0 & -1 & -1 & 0 & 0%
\end{array}%
\right) \left(
\begin{array}{c}
n_{1}^{\ast } \\
n_{2}^{\ast } \\
n_{3}^{\ast } \\
n_{4}^{\ast } \\
n_{5}^{\ast } \\
n_{6}^{\ast } \\
n_{7}^{\ast } \\
n_{8}^{\ast } \\
n_{9}^{\ast } \\
n_{10}^{\ast }%
\end{array}%
\right) =\left(
\begin{array}{c}
-a_{1} \\
-a_{2} \\
-a_{3}+\frac{D}{2} \\
-a_{4} \\
-a_{5} \\
-a_{6} \\
-a_{7}+\frac{D}{2} \\
-a_{8} \\
-a_{9} \\
-a_{10}+\frac{D}{2}%
\end{array}%
\right) .
\end{equation}%
The determinant of this matrix is $1$ and the indices $n^{\ast }$ become

\begin{equation}
\left\{
\begin{array}{l}
n_{1}^{\ast
}=3D/2-a_{1}-a_{2}-a_{3}-a_{4}-a_{5}-a_{6}-a_{7}-a_{8}-a_{9}-a_{10}, \\
n_{2}^{\ast }=-D+a_{5}+a_{6}+a_{7}+a_{8}+a_{9}+a_{10}, \\
n_{3}^{\ast }=-a_{5}, \\
n_{4}^{\ast }=-a_{6}, \\
n_{5}^{\ast }=-D/2+a_{8}+a_{9}+a_{10}, \\
n_{6}^{\ast }=-D/2+a_{1}+a_{2}+a_{3}, \\
n_{7}^{\ast }=-a_{8}, \\
n_{8}^{\ast }=-a_{9}, \\
n_{9}^{\ast }=-3D/2+a_{2}+a_{3}+a_{4}+a_{5}+a_{6}+a_{7}+a_{8}+a_{9}+a_{10},
\\
n_{10}^{\ast }=-a_{2},%
\end{array}%
\right.
\end{equation}%
that yields the value of the diagram as

\begin{equation}
\begin{array}{ll}
G= & \left( -1\right) ^{-\frac{3D}{2}}\dfrac{\Gamma \left( \frac{D}{2}%
-a_{89,10}\right) \Gamma \left( \frac{D}{2}-a_{123}\right) \Gamma \left(
\frac{3D}{2}-a_{23456789,10}\right) \Gamma \left( \frac{3D}{2}%
-a_{123456789}\right) }{\Gamma \left( a_{1}\right) \Gamma \left(
a_{4}\right) \Gamma \left( a_{7}\right) \Gamma \left( a_{10}\right) \Gamma
\left( 2D-a_{123456789,10}\right) } \\
&  \\
& \times \dfrac{\Gamma \left( a_{123456789,10}-\frac{3D}{2}\right) \Gamma
\left( D-a_{56789,10}\right) \Gamma \left( D-a_{123456}\right) \Gamma \left(
\frac{D}{2}-a_{7}\right) \Gamma \left( \frac{D}{2}-a_{4}\right) }{\Gamma
\left( D-a_{789,10}\right) \Gamma \left( D-a_{1234}\right) \Gamma \left(
\frac{3D}{2}-a_{1234567}\right) \Gamma \left( \frac{3D}{2}%
-a_{456789,10}\right) } \\
& \times \;t^{\frac{3D}{2}-a_{123456789,10}},%
\end{array}%
\end{equation}%
with the notation

\begin{equation}
a_{ijk...}=a_{i}+a_{j}+a_{k}+...
\end{equation}%
employed above.

A relevant special case is when all powers of propagators are $1$; that is, $%
a_{i}=1$ for $i=1,\cdots ,10$. This is given by

\begin{equation}
G=\left( -1\right) ^{-\frac{3D}{2}}\dfrac{\Gamma \left( 10-\frac{3D}{2}%
\right) \Gamma \left( \frac{D}{2}-3\right) ^{2}\Gamma \left( \frac{3D}{2}%
-9\right) ^{2}\Gamma \left( D-6\right) ^{2}\Gamma \left( \frac{D}{2}%
-1\right) ^{2}}{\Gamma \left( 2D-10\right) \Gamma \left( D-4\right)
^{2}\Gamma \left( \frac{3D}{2}-7\right) ^{2}}\;t^{\frac{3D}{2}-10}.
\end{equation}

\section{Conclusions}

This paper introduces a technique (GRMT) for the evaluation of a large
variety of Feynman diagrams. The advantage over previous methods is that the
evaluation of diagrams is reduced to series expansions of the integrand,
coupled with the solution of a linear system of equations.

The method is illustrated here in diagrams where the number of series
appearing in the process is the same as the dimension of the integrals
involved. Future publications will describe examples where this condition is
not present.

\bigskip

\bigskip

\textbf{Acknowledgments} The work of the second author was partially funded
by $\text{NSF-DMS }0070567$. The first and third authors also acknowledge
support from Centro Cient\'{\i}fico-Tecnol\'{o}gico de Valparaiso, CCTVal,
Chile.

\end{document}